%% file: main.tex
\documentclass[sigconf]{acmart}

\usepackage{booktabs} 
\usepackage{graphicx}
\usepackage{xcolor}
\usepackage{gensymb}
\usepackage{hyperref}
\usepackage{booktabs}
\usepackage{graphicx}
\usepackage[normalem]{ulem}
\useunder{\uline}{\ul}{}
\usepackage{enumitem}
\usepackage[absolute,overlay]{textpos}

\renewcommand\footnotetextcopyrightpermission[1]{}

\begin{document}

\begin{textblock*}{6cm}(1.5cm,26.2cm) 
	DOI: \url{https://doi.org/10.1145/3510457.3513077}
\end{textblock*}

\title{Towards a Green Quotient for Software Projects}

\author{Rohit Mehra$^\dagger$, Vibhu Saujanya Sharma$^\dagger$, Vikrant Kaulgud$^\dagger$, Sanjay Podder$^\ddagger$, Adam P. Burden*} 
\affiliation{ 
	\institution{$^\dagger$Accenture Labs, India\\ 
		$^\ddagger$Accenture, India\\
		*Accenture, USA}
}
\email{{{rohit.a.mehra, vibhu.sharma, vikrant.kaulgud, sanjay.podder, adam.p.burden}@accenture.com}}

\input{0.Abstract}


\maketitle

\input{"1.Introduction"}
\input{"2.Potential"}
\input{"3.PGQ"}
\input{"4.Conclusion"}

\bibliographystyle{ACM-Reference-Format}
\bibliography{Bibliography} 

\end{document}

%% file: 0.Abstract.tex
\begin{abstract}

As sustainability takes center stage across businesses, green and energy-efficient choices are more crucial than ever. While it is becoming increasingly evident that software and the software industry are substantial and rapidly evolving contributors to carbon emissions, there is a dearth of approaches to create actionable awareness about this during the software development lifecycle (SDLC). \textit{Can software teams comprehend how green are their projects?} Here we provide an industry perspective on why this is a challenging and worthy problem that needs to be addressed. We also outline an approach to quickly gauge the ``greenness'' of a software project based on the choices made across different SDLC dimensions and present the initial encouraging feedback this approach has received.

\end{abstract}

%% file: 1.Introduction.tex
\section{Introduction}\label{introduction}

Multiple studies have estimated that the internet and communications technology industry currently accounts for 2-7\% of the global greenhouse gas emissions, and is expected to increase to 14\% by 2040\footnote{\url{https://c2e2.unepdtu.org/wp-content/uploads/sites/3/2020/03/greenhouse-gas-emissions-in-the-ict-sector.pdf}}. These rapidly growing emissions can have a disastrous impact on the sustainability of our environment. A major share of these emissions can be attributed to the design, development, and distribution of software systems and the corresponding infrastructure required to run them. The software development process itself, and the choices/decisions that a project team makes during a typical SDLC, can have a major impact on the greenness (energy consumption and carbon emissions) of the software system.

Although there exists some body of knowledge on software sustainability, it is often overlooked and hence, its industry adoption is still limited \cite{app8112312}. For example, multiple recent studies concerning professional software developers have reported software quality, bug resolution, effective collaboration, etc. as their top priorities, but the lack of sustainability-related questions and priorities poses a major cause for concern \cite{10.1145/3368089.3409717, 7961477}. In our experience, we believe that this limited adoption can be attributed to the following challenges:

\begin{itemize}
	
	\item \textbf{\textit{Lack of Awareness}}: Despite being a first-class non-functional requirement, software sustainability is under-represented in the current software engineering curriculum \cite{10.1109/SECM.2017.4}. While the focus is predominantly on software performance, quality, security, usability, etc., software sustainability is rarely taught as an area. Moreover, we are witnessing a dearth of such courses/trainings in the open-source community as well. This acts as a major challenge for an environmentally inclined practitioner (developer, tester, program manager, etc.) who wants to adopt sustainable practices in her project, but is unaware of a starting point.

	\item \textbf{\textit{Intrusive Approaches}}: Existing approaches require in-depth access to software code, project artifacts, deployment environment, etc., to gather relevant information and suggest optimizations. This intrusive approach becomes a challenge for organizations due to adherence to multiple security, privacy, and compliance-related protocols, and hence acts as an initial barrier to adoption. The problem further amplifies for a software services organization where these artifacts are predominantly owned by the clients.

	\item \textbf{\textit{Siloed Focus Areas}}: The majority of research in this area focuses on a very specific/isolated part of the larger problem, without usually considering respective tradeoffs (impact on other parts of the project). For example, studying the effects of design patterns on energy consumption \cite{6224257}. In some cases, introducing a design pattern might increase the energy consumption of the underlying code, but removing it might have a negative impact on other software engineering dimensions like maintainability, reusability, etc. Moreover, these existing approaches tend to focus more on the technological aspects of the software project, leaving behind other project aspects such as process, team, collaboration, etc., which are equally responsible for generating emissions. This restricts the practitioner from gauging a holistic 360\textdegree \space view of the sustainability of a software project - \textit{as a whole}.

	\item \textbf{\textit{Lack of Green Metrics in Practice}}: In the current literature, there is a dearth of standardized metrics to evaluate/predict the greenness of a particular software system, especially during the design phase \cite{1d7e519992a44a2c90c95428c4ba8f9e}. It is a major challenge for the practitioners (especially architects) to proactively comprehend the impact of their design choices on the overall greenness of the software project and the to-be-developed software system. For example, how does choosing a white background color for the mobile application impacts the greenness of the software system? While there exist certain green metrics to evaluate this once the software has been developed, but in most industrial cases, it becomes an effort- and cost-intensive process to then make amends.

	\item \textbf{\textit{Hidden Impact}}: Unlike seeing carbon reductions from other activities like driving a car, it is a challenge for the practitioners to comprehend the potential benefits of a recommendation in terms of just reduction in emissions. This further limits the adoption. The eventual goal of any research in this area should be to convey the insights in a manner that highlights the potential impact in a relatable way (e.g., correlating potential emission savings with operational cost savings, or the number of rural households that can be lighted) and nudges the practitioner in adopting that recommendation (e.g., using gamification techniques).
	
\end{itemize}

In the next section, we'll discuss the potential impact on the carbon emissions of the software projects, if they can be effectively guided towards making green decisions.

\begin{figure}[t]
	\centering
	\includegraphics[width=1.0\linewidth] {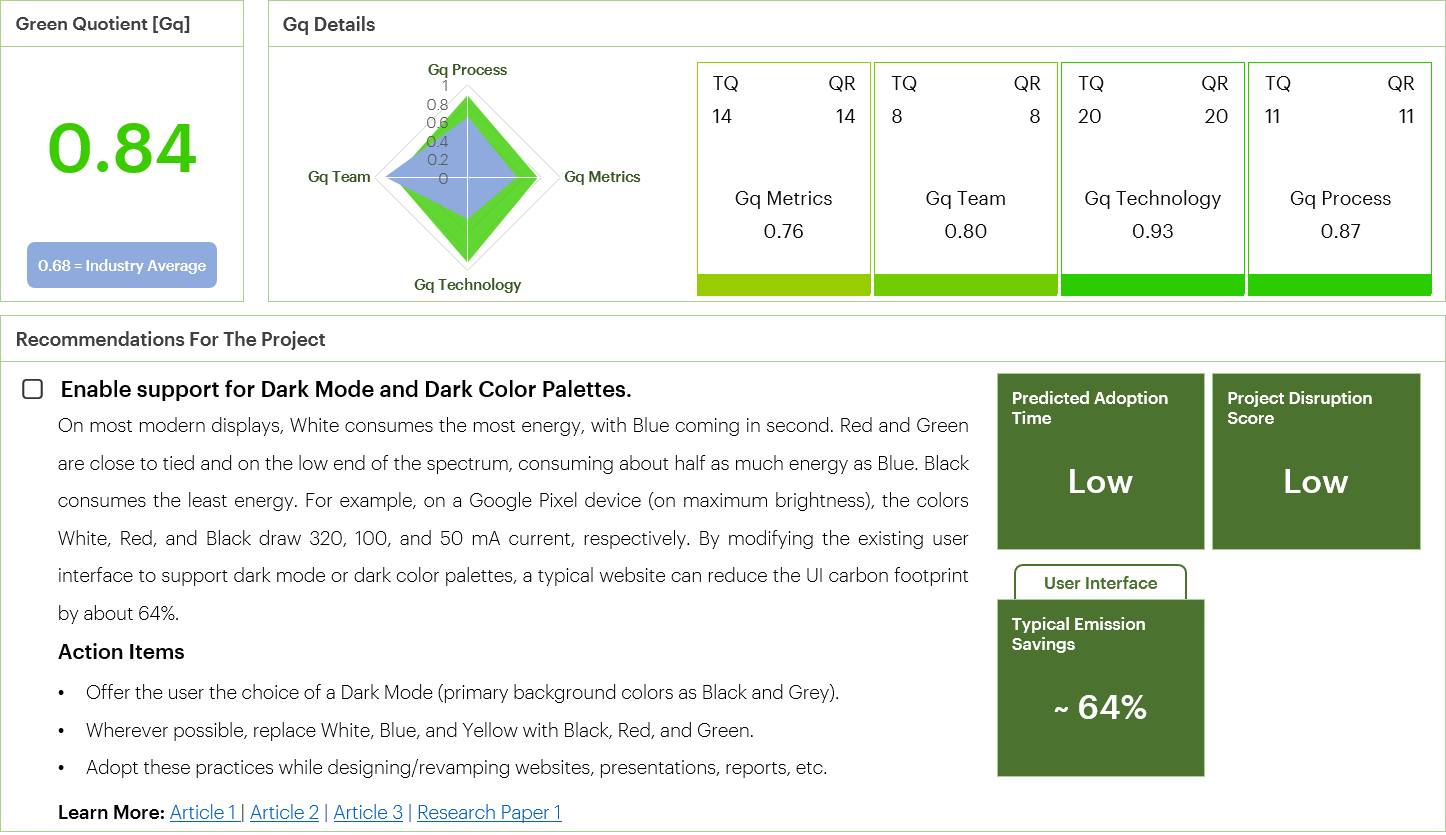}
	\caption[]{A slice of the report generated by PGQ approach.}
	\label{fig:snapshot}
\end{figure}

%% file: 2.Potential.tex
\section{Industry Potential of Making Green Decisions}\label{potential}

Throughout the SDLC, practitioners are faced with multiple decision points that guide the software development process going forward. For example, choice of programming language, choice of a third-party library, choice of a cloud provider, etc. If effectively guided, these decision points reserve the potential for making a software project - \textit{greener}. For example:

\begin{itemize}
	
	\item \textit{Programming Language}: By selecting Java over Python for developing a software system, typically about 97.4\% energy savings can be achieved \cite{10.1145/3136014.3136031}. For a large-scale software, used by millions/billions of users, this might correspond to major operational cost savings (in terms of reduced data center energy consumption), in addition to being pro-environment. Moreover, if a greener choice is not selected initially, the cost, effort, and potential emissions of code refactoring at a later stage might be exceptionally high.
	
	\item \textit{UI Color Scheme}: By choosing dark color schemes for developing a mobile application, energy consumption of the smartphone's display can be reduced by an average of 64\%, as compared to lighter color schemes \cite{10.1145/3458864.3467682}.
	
	\item \textit{Collaboration}: While collaborating remotely, by simply switching from a video meeting to an audio meeting (by turning off the camera), a typical team can reduce the carbon footprint of such meetings by about 96\% \cite{OBRINGER2021105389}. This is even more important during the current pandemic scenario when the majority of such meetings are happening remotely.
	
\end{itemize}

%% file: 3.PGQ.tex
\section{Project Green Quotient}\label{pgq}

To overcome these aforementioned challenges, we have started to explore a questionnaire-based approach, \textit{Project Green Quotient (PGQ)}, that enables a practitioner to quickly gauge the overall greenness of a software project in a non-intrusive way while receiving relevant advisory for further in-depth investigations into different non-green aspects of the project. \textit{PGQ} spans multiple project aspects such as technology, process, metrics, and team. The intent is to be the first and overarching step in a software project's journey towards being green. The approach relies on a novel metric \textit{Green Quotient}, that quantifies the overall greenness of a software project on a scale of 0 to 1 (0 representing non-adoption of any green practices and 1 representing adoption of all green practices that the approach has been trained on). Moreover, to increase adoption, the approach also highlights the potential benefits in terms of operational cost savings, emissions reduction, etc. which can be realized by adopting the system-generated recommendations. Figure \ref{fig:snapshot} showcases an early snapshot of our prototype implementation.

To understand the applicability and effectiveness of our approach, we recruited six different software projects from our delivery centers and asked them to take the PGQ assessment for their respective projects. Post completion of the assessment and generation/walkthrough of reports, the project teams were asked to complete a small qualitative feedback assessment. The results of the study are showcased in Table \ref{table:table_results} and appear very promising.

%% file: 4.Conclusion.tex
\section{Conclusion}\label{conclusion}

\input{"Tables/Table_Results"}

 Here we introduced the challenges, potential impact, and ongoing research on gaining a holistic perspective on the adoption of green decisions/practices by a software project. While the challenges and potential impact are aimed at fostering further research in this important area, the ongoing research is aimed at industry practitioners who have a desire/mandate to incorporate sustainability practices into their projects but are unaware of a starting point.

%% file: Tables/Table_Results.tex
\begin{table}[]
\scriptsize
\ttfamily
\centering
\resizebox{\columnwidth}{!}{%
\begin{tabular}{@{} p{0.01\columnwidth} p{0.84\columnwidth} p{0.01\columnwidth} p{0.07\columnwidth} p{0.07\columnwidth}  @{}}
\toprule
\textbf{}   &                                                                                                                                     & \multicolumn{3}{c}{\textbf{Feedback (\%)}}     \\ \cmidrule(l){3-5} 
\textbf{\#} & \textbf{Question}                                                                                                                   & \textbf{No} & \textbf{Probably} & \textbf{Yes} \\ \midrule
\textbf{Q1} & Did the assessment raise awareness, curiosity, and interest for further deep dive into the adoption of green practices?             &             &                   & 100          \\
\textbf{Q2} & Are these insights and recommendations helpful in incorporating green practices in the project?                                     &             & 33.33             & 66.66        \\
\textbf{Q3} & Did the assessment enable the team to think of new ideas regarding the greenness of the project?                                    &             &                   & 100          \\
\textbf{Q4} & Did the green quotient metric (and its breakdown) allow the team to gauge the overall sustainability of the project?                &             & 16.66             & 83.33        \\
\textbf{Q5} & According to you, will a similar assessment be helpful for other project teams seeking to adopt green practices in their projects?  &             &                   & 100          \\ \bottomrule
\end{tabular}%
}
\caption{Descriptive analysis of the qualitative feedback received from six early adopters of the PGQ approach.}
\label{table:table_results}
\end{table}